\input{aipcheck}
\newcommand{\met}{\ensuremath{{\slash\kern-.7emE}_{T}}}

\documentclass[
    ,final            
  ]
  {aipproc}

\layoutstyle{6x9}

\begin{document}

\title{Recent Electroweak Results from the Tevatron}

\classification{12.15.-y, 13.38.Be, 14.70.Fm, 13.85.Qk, 14.60.Cd, 13.40.Em, 14.70.Hp, 13.85.Qk}
\keywords      {Electroweak measurements, $W$ and $Z$ bosons, diboson processes}

\author{Junjie Zhu{\footnote{For the CDF and D\O~Collaborations}}}{
  address={State University of New York at Stony Brook, Stony Brook, NY, 11794}
}

\begin{abstract}
 I present the recent electroweak measurements related to single $W$, $Z$ boson
and diboson productions from the CDF and D\O~experiments at the Fermilab Tevatron 
collider.  
\end{abstract}

\maketitle


\section{Introduction}
$W$ and $Z$ bosons are mainly produced via quark-antiquark annihilations at the 
Fermilab Tevatron collider. Precision measurements with these gauge bosons provide
us with high precision tests of the Standard Model (SM) as well as indirect search
for possible new physics beyond the SM.

\section{$W$ boson properties}
\paragraph{Precision measurement of $W$ boson mass ($M_W$)} In the SM, $M_W$ can
be calculated using the electromagnetic coupling constant $\alpha$, the Fermi constant $G_F$
and the weak mixing angle $\sin^2 \theta_W$. It also receives quantum radiative corrections
that depend on the top quark mass and the SM Higgs boson mass. A precise measurement of $M_W$
thus can be used to make constaints on the Higgs mass. To make the equal 
contribution to the Higgs mass uncertainty, we need to have 
$\Delta M_W \approx 0.006 \times \Delta M_{top}$. With the current world 
average value of $\Delta M_{top}=1.3$ GeV and $\Delta M_W=0.025$ GeV, $\Delta M_W$ is 
the limiting factor for the Higgs mass constraint.

At the Tevatron, $M_W$ is extracted from a template fit to the transverse mass ($M_T$), lepton 
transverse momentum ($p_T^\ell$) and missing transverse energy ($\met$) distributions in 
$W \rightarrow \ell \nu (\ell=e, \mu)$ events. The template distributions are generated using 
a parameterized Monte Carlo (MC) simulation program. This program uses 
the state-of-the-art $W$ and $Z$ MC event generator and simulates the complex 
detector acceptance and response effects. The parameters used in the 
simulation are mainly determined using the $Z \rightarrow \ell \ell$ events. 
The uncertainties on these detector smearing parameters and theoretical 
calculations (including QCD and electroweak corrections and also the parton 
distribution functions (PDFs)) are propogated to the uncertainty on the measurement of $M_W$.

\begin{figure}
  \includegraphics[height=.25\textheight]{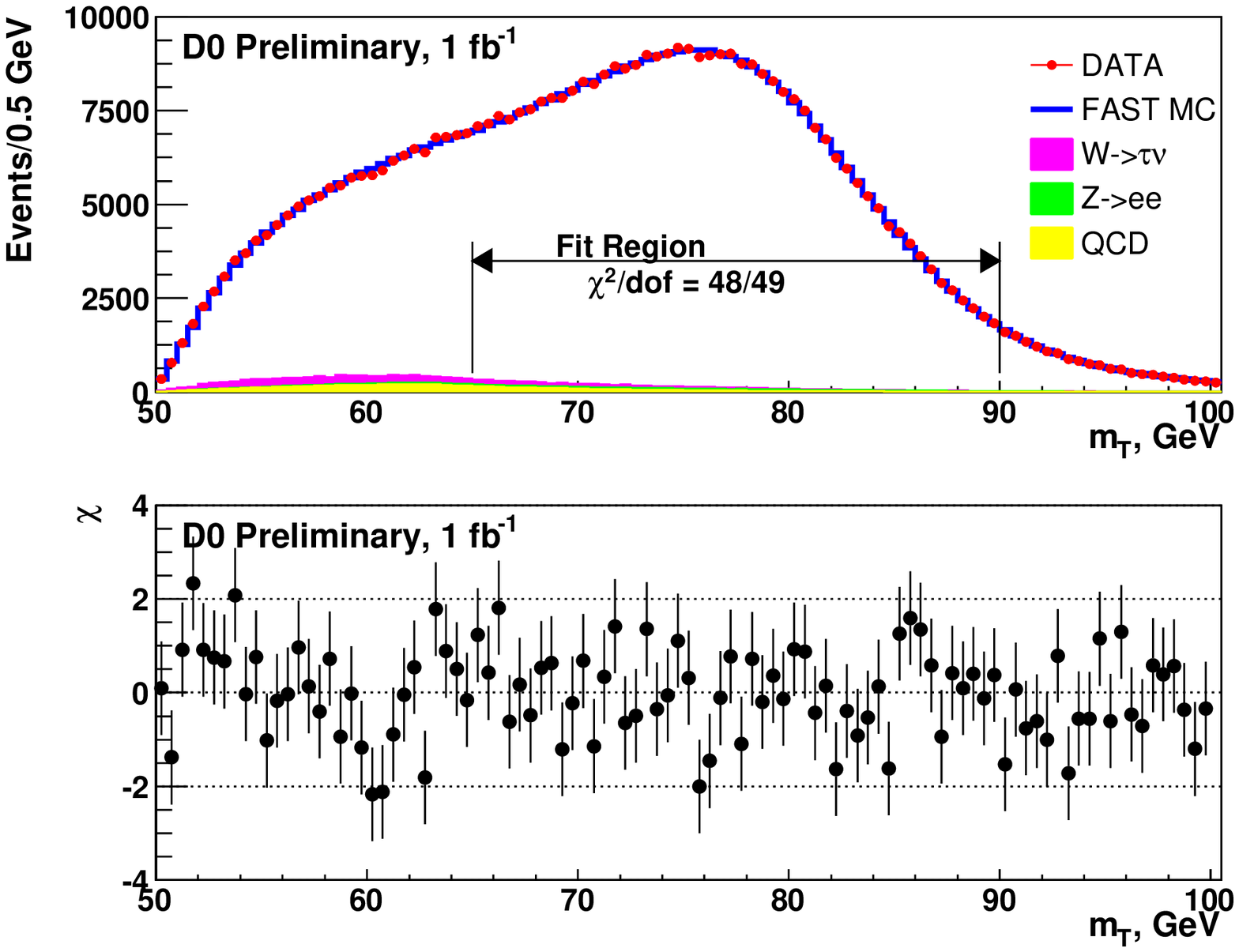}
  \includegraphics[height=.25\textheight]{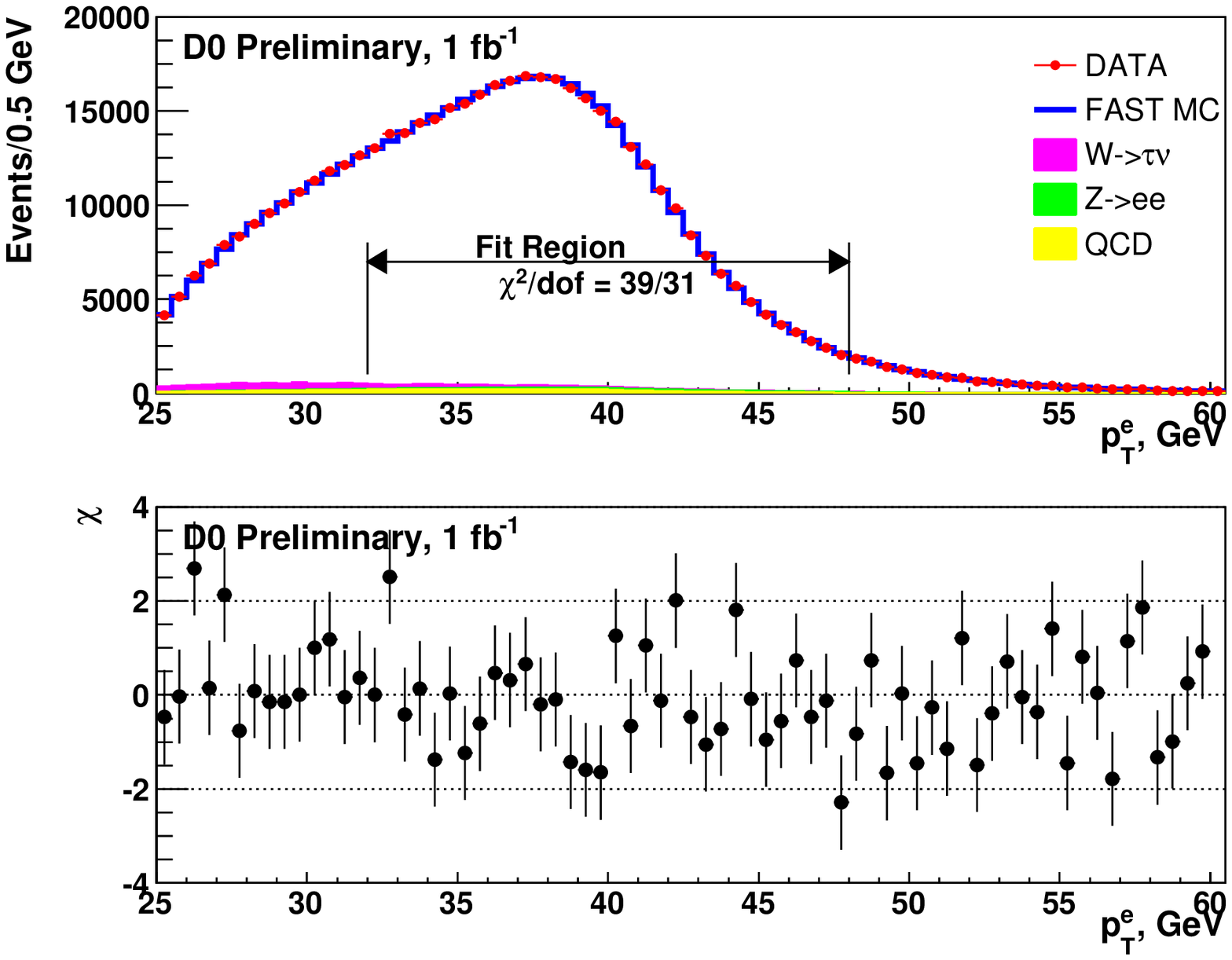}
  \caption{The $M_T$ and $p_T^e$ distributions for data and MC simulation with backgrounds added (top), 
and the difference between data and MC divided by the data uncertainty ($\chi$) for each bin (bottom).}
  \label{fig:wmass}
\end{figure}

Using 1 fb$^{-1}$ of Run II data~\cite{d0wmass}, 
the D\O~collaboration measured $M_W=80.401 \pm 0.021 {\mbox{(stat)}}\pm 0.038 {\mbox{(syst)}}= 80.401 \pm 0.043$ GeV,
the most precise measurement from one single experiment to date. Figure~\ref{fig:wmass} shows the data and 
MC comparison for $M_T$ and $p_T^e$ distributions.

Many of the systematic uncertainties in these measurements are limited by the size of the control
samples (mainly $Z \rightarrow \ell \ell$ events) used to understand the detector and 
physics effects. These uncertaines can be improved with an analysis of the 
larger datasets in hand. 
CDF has begun analyzing a data sample with 
${\cal{L}}=2.3$ fb$^{-1}$~\cite{cdfwmass} and found the statistical uncertainty is 16 (15) MeV for muon (electron)
channel and the uncertainty due to the lepton energy scale is 12 (20) MeV for muon (electron) channel. We expect 
to get the ultimate combined Tevatron measurement with an uncertainty of 15 MeV.

\paragraph{$W$ boson production charge asymmetry} At the Tevatron, $W^+$ ($W^-$) 
bosons are produced primarily by the annihilation of $u$ ($d$) quarks in 
the proton with $\bar{d}$ ($\bar{u}$) quarks in the antiproton. Due to the
fact that $u$ quarks in the proton on average carry more momentum 
than $d$ quarks, $W^+$ ($W^-$) boson tends to move along the proton (antiproton)
direction, and thus results in an asymmetry in the $W$ boson rapidity distribution between 
$W^+$ and $W^-$ boson production.
This asymmetry is sensitive to the PDFs which describes the fraction of momentum carried 
by each parton ($x$) in the proton. 

Since we can not measure the longitudinal momentum of the neutrino in 
$W \rightarrow \ell \nu$ decays, the $W$ charge asymmetry is traditionally 
measured as a function of the decay lepton pseudorapidity ($\eta$).
The D\O~collaboration analyzed 0.75 fb$^{-1}$ of data in electron channel 
and measured electron charge asymmetry for events with $E_T^e>25$ GeV, 
$p_T^{\nu}>25$ GeV and electron $|\eta|<3.2$~\cite{d0wasy}. The asymmetry is also measured
in two electron $E_T$ bins to probe partons in different $x$ regions. 
The detector effects corrected charge asymmetry is compard to a NLO perturbative
QCD calculation with CTEQ6.6 and MRST04 PDF sets. The measured charge
asymmetries tend to be lower than the theoretical predictions for high rapidity electrons. 

In a recent analysis using 1 fb$^{-1}$ of data by the CDF collaboration~\cite{cdfwasy}, the 
$W$ boson charge asymmetry is measured directly for the first time. The 
longitudinal momentum of the neutrino is estimated on an event-by-event basis 
using the $W$ boson mass constraint. The measured asymmetry is found to have 
good agreement with the predictions of a NNLO QCD calculation using the 
MRST 2006 NNLO PDF sets and a NLO QCD calculation using the CTEQ6.1 NLO PDF sets. 

For both measurements, except the largest rapidity bin ($2.8<|\eta|<3.2$) in
the D\O~measurement, the overall experimental uncertainties are smaller than 
the uncertainties due to PDFs. These measurements can be used to make tighter 
constraints on the global PDFs fits and reduce the uncertainty on the 
future $W$ boson mass measurements.

\section{Diboson Production}
The diboson production at the Tevatron is sensitive to the couplings
between gauge bosons. These triple gauge couplings (TGCs) are a direct 
consequence of the non-Abelian group structure of the SM. 
Understanding diboson production is also critical for Higgs searches where dibosons 
are a major source of backgrounds in several important channels. 
The diboson measurements at the Tevatron often involve measurements of the overall 
production cross section and TGCs. The diboson processes with one charged $W$ boson
in the final state are discussed here. The general Lorentz invariant effective 
Lagrangian describing $WWV$ ($V=\gamma$ or $Z$) vertices has five independent 
parameters with the assumption of EM gauge invariance and C and P conservation. The 
five parameters are $g^Z_1$, $\kappa_Z$, $\kappa_\gamma$, $\lambda_Z$ and $\lambda_\gamma$. 
In the SM, $g^Z_1=\kappa_{Z, \gamma}=1$ and $\lambda_{Z, \gamma}=0$. 
These couplings are often written in terms of 
their deviations from their SM values as $\Delta g^Z_1=g^Z_1-1$ and $\Delta \kappa_{Z(\gamma)}=\kappa_{Z(\gamma)}-1$. 
A dipole form factor $\alpha(\hat{s})=\alpha_0/(1+\hat{s}/\Lambda^2)^2$ 
for one arbitrary coupling $\alpha$ is introduced to avoid the unitarity violation, where $\Lambda$ is 
the new physics energy scale, and the limits are set in terms of $\alpha_0$.

\paragraph{$WW \rightarrow \ell\nu \ell\nu$} Both CDF and D\O~Collaborations select $WW$ events
with two high $p_T$ leptons and large~$\met$. Using 1 fb$^{-1}$ of data~\cite{d0ww}, D\O~observed 
22($ee)$, 64($e\mu$) and 14($\mu\mu$) candidates. These numbers are consistent with the SM predictions 
of $23.5\pm1.9$($ee$), $68.6\pm3.9$($e\mu$) and $10.8\pm0.6$($\mu\mu$) events. 
The measured cross section is $11.5\pm2.1(\mbox{stat+syst})\pm0.7(\mbox{lumi})$ pb, 
which is consistent with the SM prediction of $12.4 \pm 0.8$ pb. 
A similar measurement by CDF used 3.6 fb$^{-1}$ of data~\cite{cdfww}. 
This measurement makes use of matrix element (ME) based likelihood ratios. 
Event kinematics are used to assign an event-by-event probability ${\bf P}$ based
on leading-order ME cross section calculation for the $WW$ process and for the major background processes
such as $WW$, $W\gamma$ and $W+$jet. The background fractions are estimated using the {\sc geant} 
MC simualtion, while the signal fraction is extracted using a likelihood fit to the $LR_{WW}$ 
distribution. $LR_{WW}$ is defined as $P_{WW}/(P_{WW}+\sum_{i}k_iP_i)$ with $k_i$ as the 
relative fraction of each background source, $P_{WW}$ and $P_i$ represent the event probability 
for $WW$ and each background process respectively. 
The measured cross section is $12.1\pm0.9(\mbox{stat})^{+1.6}_{-1.4}(\mbox{syst})$ pb, which 
is again consistent with the SM prediction. 
D\O~also used the leading and trailing lepton $p_T$s to set limits on aTGCs assuming
$\Lambda=2$ TeV. The results are $-0.14<\Delta g_1^Z<0.30$, $-0.54<\Delta \kappa_{\gamma}<0.83$ and 
$0.14<\lambda_Z=\lambda_\gamma<0.19$ under the $SU(2)_L \times U(1)_Y$-conserving constraints, and 
$-0.12<\Delta \kappa_Z=\Delta \kappa_{\gamma}<0.35$, with the same $\lambda_{Z(\gamma)}$ limits as above, 
under the $WW\gamma=WWZ$ constraints.

\paragraph{$WW/WZ \rightarrow \ell\nu jj$} The D\O~Collaboration recently reported the 
first evidence of $WW/WZ \rightarrow \ell\nu jj$ process at the hadron
collider~\cite{d0wz}. The signatures are one high $p_T$ lepton, large $\met$ and two 
high $p_T$ jets. 
The background arises mainly due to $W/Z+$ jets, QCD multijet, and $t\bar{t}$
events. A multivariate classifier (Random Forest) is used to separate the signal from backgrounds. 
The signal and background contents are determined by fitting the signal and 
background random forest templates to the data. The measured cross section 
is $20.2 \pm 2.5(\mbox{stat}) \pm 3.6(\mbox{syst}) \pm 1.2(\mbox{lumi})$ pb, consistent
with the SM prediction of $16.1\pm0.9$ pb. 
The dijet $p_T$ spectrum is used to set aTGCs limits assuming $\Lambda=2$ TeV: 
$-0.12<\Delta g_1^Z<0.19$, $-0.44<\Delta \kappa_{\gamma}<0.55$ and $-0.10<\lambda_{Z, \gamma}<0.11$ under
the $SU(2)_L \times U(1)_Y$-conserving constraints, and $-0.16<\Delta \kappa_{Z, \gamma}<0.23$ 
and $-0.11<\lambda_{Z, \gamma}<0.11$, under the $WW\gamma=WWZ$ constraints.

\paragraph{{$WZ \rightarrow jj\ell\ell$}} The CDF Collaboration also reported a search for anomalous $WZ$ 
production at the Tevatron using the two charged leptons, two jets final state~\cite{cdfwz}. Three bins 
in dilepton $p_T$ are used: $105-140$ GeV (control region), $140-210$ GeV (medium region) and 
$>210$ GeV (high region). Events in the control region are mainly used to validate data modeling and 
determine backgrounds. The dijet mass $M_{jj}$ for the medium and high regions are used to set 
limits on the cross section and aTGCs. The 95\% CL upper limit on the cross section is found to 
be 234 fb and 135 fb using events in the medium and high region respectively. 
Under the $WW\gamma=WWZ$ constraints, the limits on the aTGCs are 
$-0.20<\Delta g_1^Z<0.29$, $-1.01<\Delta \kappa_{Z, \gamma}<1.27$ and 
$-0.16<\lambda_{Z,\gamma}<0.17$ assuming $\Lambda=2$ TeV, and 
$-0.22<\Delta g_1^Z<0.32$, $-1.09<\Delta \kappa_{Z, \gamma}<1.40$ and 
$-0.18<\lambda_{Z, \gamma}<0.18$ assuming $\Lambda=1.5$ TeV.

\section{Conclusions}
With the increasingly large datasets, both CDF and D\O~continue to improve our
understanding of electroweak production at the hadron collider. 

\begin{theacknowledgments}
 I would like to thank all my CDF and D\O~colleagues for their hard works,
as well as the conference organizers for hosting an excellent conference.
\end{theacknowledgments}


\IfFileExists{\jobname.bbl}{}
 {\typeout{}
  \typeout{******************************************}
  \typeout{** Please run "bibtex \jobname" to optain}
  \typeout{** the bibliography and then re-run LaTeX}
  \typeout{** twice to fix the references!}
  \typeout{******************************************}
  \typeout{}
 }


\end{document}